\documentclass[10pt,aps,prd,twocolumn,showpacs,superscriptaddress,nofootinbib,nobibnotes,floatfix]{revtex4-2}

\usepackage[utf8]{inputenc}
\usepackage[T1]{fontenc}

\usepackage{dcolumn}
\usepackage{bm}
\usepackage{mathtools,amsmath,amssymb,amsfonts,eucal,mathrsfs,graphicx,tensor,csquotes,accents,commath}
\usepackage{color}
\usepackage[dvipsnames]{xcolor}
\usepackage[unicode]{hyperref}
\hypersetup{colorlinks=true, citecolor=MidnightBlue,
            linkcolor=CornflowerBlue, urlcolor=CornflowerBlue, linktocpage=true}
\usepackage{journals}
\usepackage{wasysym}
\usepackage[normalem]{ulem}

\DeclareMathAlphabet{\mathpzc}{OT1}{pzc}{m}{it}
\definecolor{darkgreen}{rgb}{0.0, 0.6, 0.0}

\newcommand{\note}[1]{\text{\scshape\tiny{#1}}}

\newcommand{\dd}{\mathrm{d}}

\newcommand{\Ord}{\mathcal{O}}

\newcommand{\rh}{r_\mathrm{H}}

\newcommand{\al}{\alpha}
\newcommand{\be}{\beta}

\newcommand{\de}{\delta}

\newcommand{\ze}{\zeta}

\newcommand{\La}{\Lambda}

\newcommand{\om}{\omega}
\newcommand{\Om}{\Omega}

\newcommand{\bd}{\be_{\note{D},\ell}}
\newcommand{\am}{a_\note{M}}
\newcommand{\bM}{b_\note{M}}

\newcommand{\dl}{\partial}

\begin{document}

\title{Parametrized quasi-normal mode framework for non-Schwarzschild metrics}

\author{Nicola Franchini}
\affiliation{SISSA, Via Bonomea 265, 34136 Trieste, Italy and INFN Sezione di Trieste}
\affiliation{IFPU - Institute for Fundamental Physics of the Universe, Via Beirut 2, 34014 Trieste, Italy}
\affiliation{Université Paris Cit\'e, CNRS, Astroparticule et Cosmologie,  F-75013 Paris, France}
\affiliation{CNRS-UCB International Research Laboratory, Centre Pierre Binétruy, IRL2007, CPB-IN2P3, Berkeley, US}

\author{Sebastian H. V\"olkel}
\affiliation{SISSA, Via Bonomea 265, 34136 Trieste, Italy and INFN Sezione di Trieste}
\affiliation{IFPU - Institute for Fundamental Physics of the Universe, Via Beirut 2, 34014 Trieste, Italy}
\affiliation{Max Planck Institute for Gravitational Physics (Albert Einstein Institute), D-14476 Potsdam, Germany}

\begin{abstract}
In this work we comment in more detail on what happens to the parametrized framework first presented by Cardoso {\it et al}.~when there are departures from the Schwarzschild background metric, as well as possible deviations in the ``dynamics''. 
We treat possible deviations in the background metric with additional coefficients with respect to the original works. The advantages of this reformulation are clear when applied to a parameter estimation problem, since the coefficients are always real, and many of them do not depend on the overtone number and angular momentum of the frequency, thus eventually reducing the total amount of parameters to be inferred.
\end{abstract}

\maketitle

\section{Introduction}\label{introduction}

Ongoing and future efforts of gravitational wave detectors provide unprecedented access to test general relativity (GR) in the strong field regime \cite{LIGOScientific:2016lio,LIGOScientific:2018mvr,LIGOScientific:2020ibl,LIGOScientific:2021djp}. 
Among the several exciting possibilities, one is to perform black hole spectroscopy in order to verify the validity of the requirements of the no-hair theorem and also general relativity itself \cite{PhysRevLett.11.237,PhysRevLett.26.331,PhysRevLett.34.905}. Our current understanding of binary black hole mergers predicts that the final stage of such events should be described by a superposition of linear perturbations of the background, the so-called quasi-normal modes~\cite{Regge:1957td,Zerilli:1970se,Teukolsky:1973ha}, see also Refs.~\cite{Kokkotas:1999bd,Nollert:1999ji,Berti:2009kk,Konoplya:2011qq} for reviews. 

The spectrum of quasi-normal modes is sensitive to the background perturbed metric, but also to the dynamical behaviour of the theory in which the perturbations are excited. The purpose of this work is to provide a framework which allows one to study quasi-normal modes in spherical symmetry for metrics which are different from the Schwarzschild metric when the modifications are small. 
Our framework follows a prescription similar to the one presented by Cardoso et \textit{al.}~\cite{Cardoso:2019mqo} and McManus et \textit{al.}~\cite{McManus:2019ulj}. 
In these works, a modified (system) of wave equations has been introduced and the corresponding quasi-normal mode spectrum has been computed up to quadratic order for a set of small deviation parameters. 
We explicitly show how possible deviations in the background metric could affect the quasi-normal frequencies, in a different fashion compared to~\cite{Cardoso:2019mqo,McManus:2019ulj}. 
While the general idea is similar, our approach allows for a better description of non-Schwarzschild backgrounds, purely real deviation parameters in the framework, as well as a differentiation between the parameters that depend on the angular momentum of the frequencies and those which do not.

This work is structured as follows. 
In Sec.~\ref{theoretical_minimum} we introduce the metric dependent parametrized framework. We apply it to two different cases in Sec.~\ref{app}. 
Conclusions are discussed in Sec.~\ref{conclusions}. 
Throughout this work we use units in which $G = c = 1$.

\section{Theoretical Framework}\label{theoretical_minimum}

Let us start by qualitatively reviewing how gravitational perturbations in gravity theories can be studied. In general, one assumes a linear perturbation of the metric as
\begin{equation}
    g_{ab} \simeq \overline{g}_{ab} + h_{ab}\,,
\end{equation}
for which the background metric $\overline{g}_{ab}$ and the perturbation metric $h_{ab}$ satisfy the 0th order and 1st order equations of motion, respectively. They are obtained from the linear expansion of the full system of equations
\begin{equation}
    G_{ab} \simeq \overline{G}_{ab} + H_{ab} = 0\,,
\end{equation}
where $G_{ab}$ are the equations of motion of a given gravity theory.
Let us assume that the background metric is given by some non-GR metric. We remark that we do not know {\it a priori} which equations of motion this metric is satisfying. However, as long as it is perturbatively close, with respect to some parameter $\de$, to the Schwarzschild/Kerr metric, it must satisfy Einstein's equations of GR up to some error linear in $\de$,
\begin{equation}
    \overline{G}^\note{GR}_{ab} \left[\overline{g}_{cd}\right] = 0 + \Ord\left(  \de \right) \,.
\end{equation}
The presence of the $\Ord(\de)$ term can be interpreted as the fact that the equations of motion of a new theory whose static solution is approximated by the metric $\overline{g}_{ab}$ must have different dynamics, not captured with the GR equations by a linear factor.

We can conjecture that the first order equations share the same property. For GR, one would have
\begin{equation}\label{eq:1st_GR}
    H_{ab}^\note{GR} \left[ h_{cd} ; \, \overline{g}_{cd}^\note{Sch} \right] = 0 \,,
\end{equation}
but when $\overline{g}_{cd}^\note{Sch}$ is substituted by $\overline{g}_{cd}$, the equations become
\begin{equation}\label{eq:1st_mod}
     H_{ab}^\note{GR} \left[ h_{cd} ; \, \overline{g}_{cd}\right] = 0 + \Ord(\de) \,.
\end{equation}
One must bear in mind that this notation generically includes the 0th and 1st order metric as well as their derivatives.

Let us now take into account a spherically symmetric background metric defined as follows
\begin{equation}\label{eq:metric_mod}
    \overline{g}_{ab} \dd x^a \dd x^b = -f(r) A(r)\dd t^2 + \frac{B^2(r)}{f(r) A(r)}\dd r^2 + r^2 \dd\Om\,,
\end{equation}
with $f(r) = 1 - \rh/r$, $\rh$ being the location of the event horizon.
Then, we can expand the functions $A$ and $B$ with a set of {\it real} coefficients $\am^{(k)}, \bM^{(k)}$ in a post-Newtonian fashion~\cite{Barausse:2014tra}
\begin{align}
A(r) & = 1+ \de\sum_{k=1}^K \am^{(k)}\left(\frac{\rh}{r}\right)^k \,, \\
B(r) & = 1+ \de\sum_{k=1}^K \bM^{(k)}\left(\frac{\rh}{r}\right)^k \,.
\end{align}
We make the series terminate at some finite number $K$ starting from $k=1$ to ensure asymptotic flatness.

From the prescription suggested by passing from Eq.~\eqref{eq:1st_GR} to Eq.~\eqref{eq:1st_mod} one could guess the following form for the master perturbation equation
\begin{equation}\label{eq:master_intermediate}
    f Z \frac{\dd}{\dd r} \left( f Z \frac{\dd \Phi}{\dd r} \right) + \left[ \om^2 W - f Z \overline{V}_\ell \right] \Phi = 0\,,
\end{equation}
where one has enough generality to define the perturbation variable $\Phi$ to choose $Z=A/B$ such that the tortoise coordinate $r_*$ arising from this equation is $\dd r_* \, f A/B = \dd r$.
Finally, we expect small deviations from GR in the other parameters as well: $W = 1 + \de w(r)$, $\overline{V}_\ell = \overline{V}^\note{GR}_\ell(r) + \de \overline{V}_\ell(r)$. Note that in this definition we made explicit that radial functions modifying the potential depend on the angular momentum $\ell$ of the quasi-normal frequencies. 

In the following we try to provide a heuristic explanation for why we expect this form for the master equation.
The first step is to write down the master equation for a perturbation in GR
\begin{equation}\label{eq:master_GR}
F_\note{GR} \frac{\dd}{\dd r}\left( F_\note{GR} \frac{\dd {\Phi}}{\dd r} \right) + \left[ \om^2 - \overline{V}^\note{GR} \right]{\Phi} = 0\,,
\end{equation}
where $F_\note{GR} \equiv \sqrt{\overline{g}^\note{Sch}_{tt} \, \overline{g}^{\,rr}_\note{Sch}} = f(r)$ is the ``natural'' tortoise function for the Schwarzschild solution, and
\begin{equation}
    \overline{V}^\note{GR} = \overline{g}^\note{Sch}_{tt} \frac{\La}{r^2} + G(r)\frac{F_\note{GR}}{r} \frac{\dd F_\note{GR}}{\dd r}\,,
\end{equation}
where $\La = \ell(\ell+1)$, $G(r)=1$ denotes scalar perturbations, $G(r)=-3$ for axial perturbations and
\begin{equation}\label{eq:coeff_polar}
    G(r) = -3\frac{1 + 3F_\note{GR}^2(r) - \La^2}{\left[1 -3F_\note{GR}(r) + \La \right]^2}
\end{equation}
for polar perturbations. 
We intentionally wrote the eikonal term $\overline{g}^\note{Sch}_{tt}\La/r^2$ to make its correspondence with geodesic motion evident.

If we assume that the background metric is the metric of Eq.~\eqref{eq:metric_mod}, we can substitute all the $\overline{g}_{ab}^\note{Sch}$ terms with $\overline{g}_{ab}$ everywhere in equation~\eqref{eq:master_GR}. For example, one should transform $F_\note{GR}$ into $f A/B$. Moreover, to take into account that the theory can be different, one should add a correction $\de Z$ to the tortoise coordinate and one $\de \overline{V}$ to the potential. These two corrections take into account the fact that the derivation of the perturbation equation might be done in a theory which is not GR [{\it cf.}~Eq.~\eqref{eq:1st_mod}]. The resulting equation is
\begin{equation}
\begin{multlined}
    f (Z + \de Z) \frac{\dd}{\dd r} \left[ f (Z + \de Z) \frac{\dd \Phi}{\dd r} \right] \\
    + \left[ \om^2 - f Z \left(V^\note{GR}_\ell + \de \overline{V}_\ell \right) \right] \Phi = 0\,.
\end{multlined}
\end{equation}
One can always re-define the perturbation function $\Phi$ and re-scale the equation to remove the $\de Z$ term from the derivative term. This would ensure that the second order derivative is acted upon the tortoise coordinate of the metric. However, by doing so, one would introduce some terms multiplying the frequency, obtaining Eq.~\eqref{eq:master_intermediate}. The term multiplying the frequency can be mapped into the modification of the tortoise coordinate $\de w(r) = -2\de Z(r)$, and we expand it as
\begin{equation}
    \de w(r) = \sum_{k=1}^K w^{(k)} \left(\frac{\rh}{r} \right)^k \,.
\end{equation}
It is worth noting that this term appears because the background metric (and thus the tortoise coordinate) is different from the Schwarzschild metric.

On the potential side, from the construction, we can split the contribution to the potential coming from the metric and from the ``dynamics''. We have that 
\begin{align}
\overline{V}^\note{GR}_\ell + \de \overline{V}_\ell = \overline{V}^\note{mod}_\ell + \de \overline{V}_\ell^\note{D},
\end{align}
where we made the following identifications:
\begin{align}
    \overline{V}_\ell^\note{mod} & = B\frac{\ell (\ell+1)}{r^2} + \frac{G(r)}{r} \frac{\dd }{\dd r} \left( f \frac{A}{B} \right)\,, \\
    \de \overline{V}_\ell^\note{D} & = \frac{\de}{\rh^2}\sum_{k = 0}^K \bd^{(k)}\left(\frac{\rh}{r}\right)^k \,.
\end{align}
This last term, linear in $\de$, expresses in a $1/r$ basis our ignorance on the correct equations of motion for the dynamics of the theories under scrutiny.

Collectively we label all the {\it real} coefficients, from metric and from dynamics as
\begin{equation}
    \be^{(i)} = \left( \bd^{(0)}, z^{(1)}, \bM^{(1)}, w^{(1)}, \bd^{(1)}, z^{(2)}, \bM^{(2)}, w^{(2)}, \bd^{(2)}, \dots \right)\,,
\end{equation}
where we identified $z^{(k)} = \am^{(k)} - \bM^{(k)}$ from the definition of $Z$. In the rest of the paper we will refer to this set of parameters as {\it mixed} parametrization or {\it metric-potential} parametrization to distinguish it from the {\it potential-only} parametrization introduced in~\cite{Cardoso:2019mqo}.
A quadratic expansion of $\om$ in the parameters $\be^{(i)}$ is simply given by
\begin{equation}
    \om \simeq \om^0 + c_{(i)}\be^{(i)} + \frac{1}{2} v_{(ij)}\be^{(i)}\be^{(j)}\,.
\end{equation}

We can infer the coefficients $c_{(i)}$ and $v_{(ij)}$ from the coefficients $d_{(k)}$ and $e_{(ks)}$ obtained in~\cite{Cardoso:2019mqo,McManus:2019ulj}\footnote{See also~\cite{Kimura:2020mrh,Volkel:2022aca} for alternative ways to compute these coefficients} by comparing the respective master equations.
In order to do so, we must manipulate Eq.~\eqref{eq:master_intermediate}. By performing the change of variables $\phi = \sqrt{Z}\Phi$, the master equation takes the form
\begin{gather}
    \label{eq:master_temp}
    f \frac{\dd}{\dd r}\left[ f \frac{\dd \phi}{\dd r} \right] + \left[ \om^2\frac{W_\mathrm{H}}{Z_\mathrm{H}^2} - f V \right]\phi = 0 \,, \\
    V = \frac{\overline{V}}{Z} - \frac{f \left(Z'\right)^2 - 2 Z \left( f Z' \right)'}{4 Z^2}  - \frac{\om^2}{f}\left( \frac{W}{Z^2} - \frac{W_\mathrm{H}}{Z_\mathrm{H}^2}\right) \label{eq:Vtemp}\,,
\end{gather}
with $Z_\mathrm{H} = Z(\rh)$ and $W_{\mathrm{H}}=W(\rh)$. In addition, the last term of \eqref{eq:Vtemp} can be put in the $1/r$ form with the following manipulation:
\begin{align}
    \frac{\om^2}{f}\Bigg( \frac{W}{Z^2} \,-\, & \frac{W_\mathrm{H}}{Z_\mathrm{H}^2}\Bigg) \notag = \om^2 \sum_k \frac{2z^{(k)} - w^{(k)}}{r^{k-1}} \frac{r^k - \rh^k}{r - \rh} \notag \\
    & = \om^2 \sum_k \left(2z^{(k)} - w^{(k)} \right) \sum_{p=0}^{k-1} \left(\frac{\rh}{r}\right)^p \,.
\end{align}
In this way, we can define a first order expansion of the potential in $\de$ as
\begin{equation}\label{eq_mod_V}
    \de V \equiv V - V_\note{GR} = \frac{\de}{\rh^2}\sum_{k=0}^\infty \al^{(k)}\left( \frac{\rh}{r} \right)^k \,.
\end{equation}
The coefficients $\al^{(k)}$ can be obtained up to second-order by combinations of the coefficients $\be^{(i)}$. 
For generic index $k$ and tensor axial perturbation, we have that linear contributions are
\begin{align}
    \al^{(k)} = & \, \left[ (k-2) z^{(k-2)} - (k-3) z^{(k-3)} \right] \frac{k-7}{2} \notag \\
    & + \La \left( \bM^{(k-2)} - z^{(k-2)} \right) + \bd^{(k)} \notag \\
    & + \om^2\rh^2 \sum_{p=k+1}^K \left(w^{(p)} - 2z^{(p)} \right) \,,
\end{align}
while we do not display the quadratic ones since the expression is lengthy and uninformative. 
The formula for scalar perturbations has an analogous form, while for the tensor polar case, one should first decompose the term~\eqref{eq:coeff_polar} such that it yields $1/r$ terms. 
We suggest that the best-motivated way to proceed is to perform an eikonal expansion of the term ({\it i.e.}, around $\ell\rightarrow\infty$) and include eikonal terms until the deviation in the coefficients is smaller than the numerical error of the coefficients themselves.

The manipulation shown with the previous calculations results in equation~\eqref{eq:master_temp} being the same form as that in Ref.~\cite{Cardoso:2019mqo}. 
Now, one can apply the quadratic expansion of the quasi-normal frequencies from Refs.~\cite{Cardoso:2019mqo,McManus:2019ulj} to Eq.~\eqref{eq:master_temp}, yielding
\begin{equation}
\begin{multlined}
    \om \simeq \frac{Z_\mathrm{H}}{W_\mathrm{H}^{1/2}}\Bigg(\om^0 + d_{(k)} \al^{(k)} \\ + d_{(k)}d_{(s)}\al^{(k)}\dl_\om\al^{(s)} + \frac{1}{2}e_{(ks)}\al^{(k)}\al^{(s)}\Bigg)\,.
\end{multlined}
\end{equation}
Due to the relation between the coefficients $\al^{(k)}$ and $\be^{(i)}$, we can always do the mapping
\begin{align}
    c_{(i)} \equiv & \, \frac{\dl \om}{\dl\be^{(i)}} =  \om_0\frac{\dl }{\dl\be^{(i)}}\frac{Z_\mathrm{H}}{W_\mathrm{H}^{1/2}} + d_{(k)} \frac{\dl\al^{(k)}}{\dl\be^{(i)}} \,, \\
    v_{(ij)} \equiv & \, \frac{\dl^2\om}{\dl\be^{(i)}\dl\be^{(j)}} = \om_0\frac{\dl^2}{\dl\be^{(i)}\dl\be^{(j)}}\frac{Z_\mathrm{H}}{W_\mathrm{H}^{1/2}} \notag \\
    & + d_{(k)} \left( \frac{\dl}{\dl\be^{(i)}}\frac{Z_\mathrm{H}}{W_\mathrm{H}^{1/2}} \frac{\dl\al^{(k)}}{\dl\be^{(j)}} +\frac{\dl}{\dl\be^{(j)}}\frac{Z_\mathrm{H}}{W_\mathrm{H}^{1/2}} \frac{\dl\al^{(k)}}{\dl\be^{(i)}} 
    \right) \notag \\
    & + d_{(k)} d_{(s)} \left( \frac{\dl\al^{(k)}}{\dl\be^{(i)}} \frac{\dl^2\al^{(s)}}{\dl\be^{(j)}\dl\om} + \frac{\dl\al^{(k)}}{\dl\be^{(j)}} \frac{\dl^2\al^{(s)}}{\dl\be^{(i)}\dl\om}  \right)\notag \\
    & + d_{(k)} \frac{\dl^2\al^{(k)}}{\dl\be^{(i)}\dl\be^{(j)}} + e_{(ks)} \frac{\dl\al^{(k)}}{\dl\be^{(i)}} \frac{\dl\al^{(s)}}{\dl\be^{(j)}}.
\end{align}

It is worth noting that since all the coefficients $\be^{(i)}$ are real, any term that would make the coefficients $\al^{(k)}$ complex comes from the fact that the background metric is not Schwarzschild. 
Indeed, we note that the coefficients $w^{(k)}$ denote the deviation of the ``natural'' tortoise coordinate of the master equation from the tortoise coordinate of the metric, nonetheless defined by the coefficients $z^{(k)}$.\footnote{An additional frequency-dependent contribution not shown here could come from a non-spherically symmetric background metric, that reduces to Schwarzschild in the zero coupling limit ({\it e.g.}, the slow-rotation expansion~\cite{Pani:2013pma}).} 

The advantage of this formulation over the potential-only parametrization is evident when reconstructing the modified potential after the detection of more than one quasi-normal mode~\cite{Volkel:2020daa,Volkel:2022aca,Volkel:2022khh}. 
In fact, even at first it seems that the metric-potential formalism has more parameters than the potential-only one, one must be aware that the coefficients of the latter are complex (because they can depend on the frequency) and they are different for each mode measured, as they are sensitive to $n$ and $\ell$. This means twice as many $K+1$ coefficients for each mode measured.
On the other hand, the mixed formalism has $K+1$ coefficients $\bd^{(k)}$ which change with the angular momentum of the mode, and $K-1$ coefficients $z^{(k)}$, $\bM^{(k)}$ and $w^{(k)}$. None of them depends on the overtone number. 
By simple counting, one can find that the total number of coefficients for the observation of $n_\ell$ different angular momentum modes and $n_n$ additional ({\it i.e.}, on top of the fundamental one) overtones for the potential-only case is $2(n_n + n_\ell)(K+1)$, while for the metric-potential case, one has $n_\ell(K+1) + 3(K-1)$. 

Formally the mapping to the metric can also be extended to cover the more general scenario of a coupled system of equations, as studied in Ref.~\cite{McManus:2019ulj}, if one makes similar assumptions about how the fields are coupled to each other.
Those key assumptions are that the additional coupling functions can be expressed in a similar expansion scheme as the potential in eq.~\eqref{eq_mod_V} and that the additional fields couple linearly via the coupling functions. 
Such a structure of the coupled case has been explored in Ref.~\cite{McManus:2019ulj} for dynamical Chern-Simons, Horndeski gravity and an effective field theory approach. 

\section{Applications}\label{app}

\subsection{Reissner-Nordstr\"om metric}\label{app_RN}

\begin{figure}[h!]
\centering
\includegraphics[width=\linewidth]{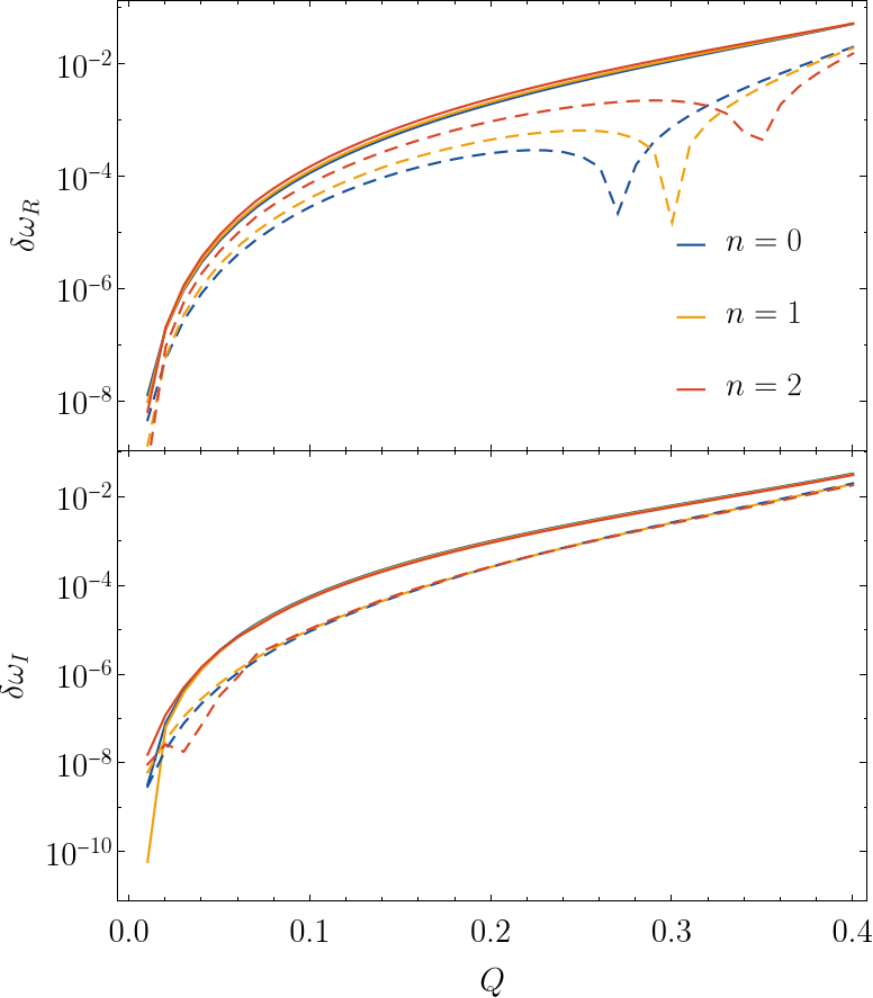}
\caption{Relative errors $\de \om = |\om-\om^\note{RN}|/\om^\note{RN}$ for the real (upper panel) and imaginary (lower panel) parts of the axial $\ell=2$ RN quasi-normal modes for $n=0,1,2$ using the potential parametrized framework (solid lines) and the new metric dependent one (dashed lines).}
\label{fig_RN}
\end{figure}

As first non-trivial application of the framework we consider the Reissner-Nordstr\"om (RN) metric \cite{https://doi.org/10.1002/andp.19163550905,https://doi.org/10.1002/andp.19173591804,1918KNAB...20.1238N,jeffery1921field}. With the current notation it is given by the metric Eq.~\eqref{eq:metric_mod}, with
\begin{align}
A = 1 - \frac{r_{-}}{r}, \qquad B = 1,
\end{align}
where $r_-$ can be related to the mass $M$ and a (small) electric charge $Q$ through
\begin{align}
r_{-} = M - \sqrt{M^2-Q^2} \approx \frac{Q^2}{2 M},
\end{align}
and the location of the horizon is given by
\begin{align}
r_\text{H} = M + \sqrt{M^2 - Q^2}.
\end{align}
The perturbation equation in the odd-parity sector is close to the Regge-Wheeler equation~\cite{Berti:2009kk}. From its expression, one can either infer the axial deviation parameters for potential-only formalism ({\it cfr.}~Ref.~\cite{Cardoso:2019mqo})
or the parameters for the mixed framework
\begin{align}
    z^{(1)} & = -\frac{r_{-}}{\rh}\,, \\
    \bd^{(3)} & = -\frac{4r_{-}}{3\rh}(\La-2)\,, \qquad \bd^{(4)} = -2\frac{r_{-}}{\rh} \,.
\end{align}
In Fig.~\ref{fig_RN} we compare the application of the original parametrized framework with the new metric dependent one for the RN $n=0,1,2$, $\ell=2$ quasi-normal modes. 
In the two panels we show the relative error $\de\om = |\om - \om^\note{RN}|/\om^\note{RN}$ between the parametrized quasi-normal modes $\om$ (potential-only parametrization is represented by solid lines while potential-metric parametrization by dashed lines) and the numerically computed one $\om^\note{RN}$. We check that from the scaling of the relative error, the deviation of RN QNMs from Schwarzschild QNMs is much larger than the error introduced by the parametrization, for the parameters considered.
One can clearly see that the mixed parametrization gives an even more accurate prediction of the frequencies for each case considered. 
We also checked different angular momentum $\ell$ which gives analogous features.

\subsection{Einstein-dilaton-Gauss-Bonnet gravity}

Einstein-dilaton-Gauss-Bonnet (EdGB) gravity is a well-studied alternative theory of gravity, where a dynamical scalar field is coupled to the Gauss-Bonnet invariant through a coupling $\al_\note{GB}$. Our notation for the action of the theory and the equations of motion is consistent with that of Ref.~\cite{Bryant:2021xdh}. Further information about the theory can be found in the latter, as well as in Ref.~\cite{Julie:2019sab} and references therein.
In the small-coupling limit, unless the coupling function does not have a linear term in the scalar field, all BHs in the theory develop a scalar profile around them, and the background metric and the perturbation equations are of the form we analysed here. The QNM spectrum of these solutions was computed in Ref.~\cite{Blazquez-Salcedo:2016enn} without the small coupling expansion. The master equation for axial perturbation can be recast in the form of Eq.~\eqref{eq:master_intermediate} retaining up to $\ze^2$ terms, where we introduced the dimensionless coupling $\ze = 4\al_\note{GB}/\rh^2$. The relevant functions are (according to the notation of~\cite{Bryant:2021xdh})
\begin{align}
    B & = 1 - \ze^2 \left( \frac{\rh^2}{8r^2} + \frac{\rh^3}{6r^3} + \frac{7\rh^4}{16r^4} + \frac{2\rh^5}{5r^5} + \frac{3\rh^6}{8r^6} \right) \,, \\
    Z & = 1 + \ze ^2 \Bigg(-\frac{49 \rh}{80 r}-\frac{39\rh^2}{80 r^2}-\frac{97 \rh^3}{240 r^3} \notag \\
    & \qquad\qquad\quad +\frac{49 \rh^4}{120 r^4}+\frac{61 \rh^5}{120r^5}+\frac{7 \rh^6}{12 r^6}\Bigg) \,, \\
    W & = 1+ \ze ^2 \left(-\frac{2\rh^3}{r^3} -\frac{\rh^4}{r^4} -\frac{\rh^5}{r^5} +\frac{4
   \rh^6}{r^6} \right) \,,
\end{align}
\begin{align}
   \overline{V}_\ell & = V^\note{GR}_\ell + \ze^2 \frac{ \rh}{r^3} \Bigg[
   -\frac{147}{80}
   +\frac{(6-\Lambda )\rh}{8 r} \notag \\
   &+\left(\frac{99}{4}-\frac{19 \Lambda }{6}\right)\frac{\rh^2}{r^2}
   +\left(-\frac{15 \Lambda }{16}-\frac{29}{4}\right)\frac{\rh^3}{r^3} \notag  \\
   &-\frac{9 (\Lambda -5) \rh^4}{10r^4}
   +\left(\frac{41 \Lambda }{8}-\frac{712}{5}\right)\frac{\rh^5}{r^5}
   +\frac{519 \rh^6}{4 r^6}
   \Bigg] \,.
\end{align}

\begin{figure}
\centering
\includegraphics[width=\linewidth]{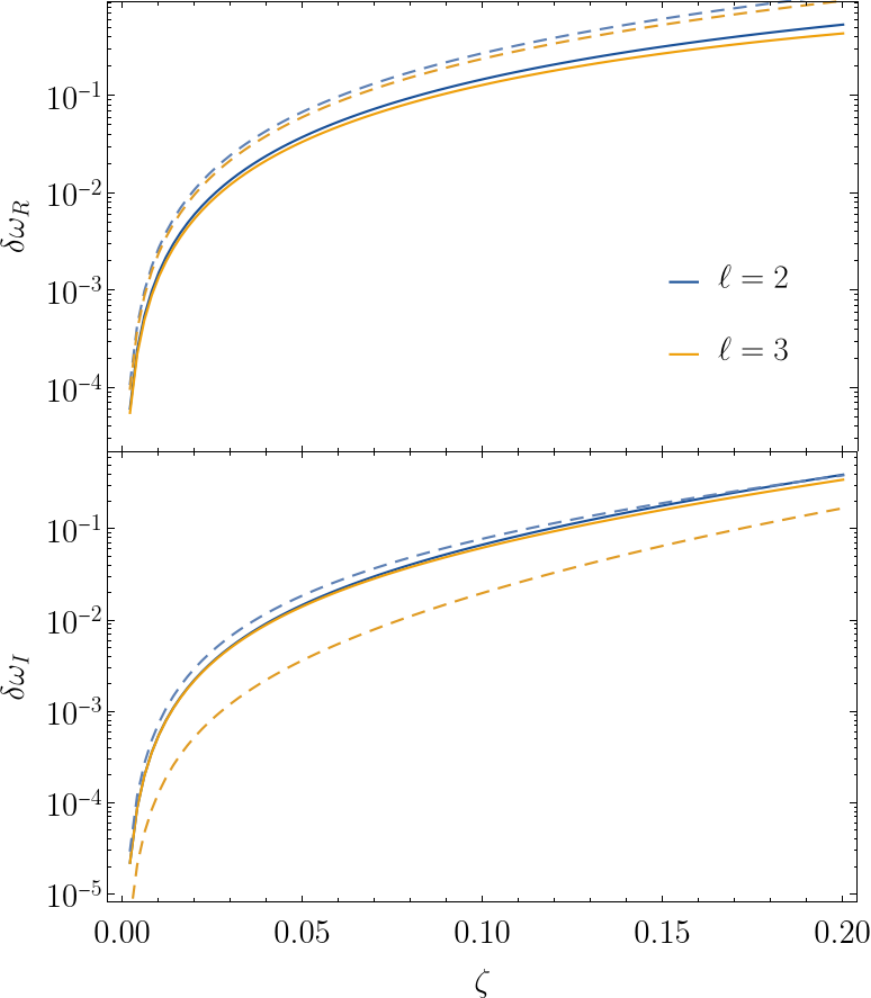}
\caption{Relative errors $\de \om = |\om-\om^\note{GB}|/\om^\note{GB}$ for the real and imaginary parts of the axial $n=0$ GB quasi-normal modes for $\ell = 2,3$ using the potential-only parametrized framework (solid lines) and the metric-potential dependent one (dashed lines).}
\label{fig_GB}
\end{figure}

As shown in section~\ref{theoretical_minimum}, one can transform the equation such that it is in the form suitable to express the modifications from GR as coefficients of the potential formalism. 
Hence, in Fig.~\ref{fig_GB}, we compare the $\ell=2,3$ fundamental modes obtained with the two different formalisms against the fits of the numerical quasi-normal mode frequencies $\om^\note{GB}$ computed in~\cite{Blazquez-Salcedo:2016enn}. 
In this case, the metric-potential parametrization under-performs the potential-only one in most cases, even though the error with respect to the numerical fits are still comparable.

\section{Conclusions}\label{conclusions}

In this work we have introduced a framework to compute quasi-normal modes from non-GR metrics where modifications are small. 
Our ansatz is very similar and can be mapped to the one proposed by Cardoso et \textit{al.}~\cite{Cardoso:2019mqo}. This allows for a quick computation of QNMs of a given metric. We have found comparable precision among the two methods in approximating the numerically computed frequencies both for RN and EdGB axial perturbations.
We showed how to perform the mapping between the two formalisms for scalar and tensor axial perturbations only, but we also provided an explanation on how to do it for the tensor polar case.

The main advantage of the mixed method is evident when one has in mind the so-called inverse problem, which refers to the reconstruction of the perturbation equation in a theory-agnostic approach~\cite{Volkel:2020daa,Volkel:2022aca,Volkel:2022khh}. The essential condition for the inverse problem is the detection of more angular modes/overtones simultaneously, even if the capability of doing so with current observations resulted in an ongoing discussion in the literature~\cite{Isi:2019aib,Capano:2021etf,Cotesta:2022pci,Finch:2022ynt,Isi:2022mhy,Capano:2022zqm}. The advantage in the inverse problem comes from a reduction of the number of parameters, all real, and only with the $\bd^{(k)}$ depending on the angular momentum of the perturbation. 
This is not the case for the potential-only parametrization, for which all coefficients $\al^{(k)}$ are, in principle, complex and depend on angular momentum and overtone number as well. Moreover, it can give insight into discerning whether a modification comes from the background or from the dynamics.

The possibility to explicitly separate the two contributions could also be useful when complementary constraints on the metric are provided, but not on the dynamics. 
Such a case may be possible with activities of the Event Horizon Telescope Collaboration \cite{EventHorizonTelescope:2019dse,EventHorizonTelescope:2019ggy,EventHorizonTelescope:2022wkp}, which can, in principle, be used to study possible deviations from the Schwarzschild/Kerr metric of supermassive black holes, see e.g. Refs.~\cite{EventHorizonTelescope:2020qrl,Volkel:2020xlc,EventHorizonTelescope:2021dqv,Nampalliwar:2021oqr,Lara:2021zth,Kocherlakota:2022jnz,Ayzenberg:2022twz,EventHorizonTelescope:2022xqj} for some recent works. 
Other complementary constraints can also be obtained using x-ray spectroscopy; see, e.g., Refs.~\cite{Bambi:2016sac,Tripathi:2018lhx,Cardenas-Avendano:2019zxd}. 
Although it seems unlikely that observations of the same objects can be done using gravitational waves and one of these techniques in the foreseeable future, one can still combine constraints on individual metrics under certain assumptions, {\it e.g.}, assuming that the way the metric is modified is similar for all sources. 

\acknowledgments
We thank Enrico Barausse, Andrea Maselli and Masashi Kimura for providing feedback on the final version of the manuscript.
We acknowledge financial support provided under the European Union's H2020 ERC Consolidator Grant ``GRavity from Astrophysical to Microscopic Scales'' grant agreement no.~GRAMS-815673. S.V.~acknowledges funding from the Deutsche Forschungsgemeinschaft (DFG) - project number: 386119226.

\bibliography{literature}

\end{document}